\documentclass[pre,twocolumn]{revtex4-2}
\usepackage[utf8]{inputenc}
\usepackage{amsmath}
\usepackage{amsfonts}
\usepackage{amsthm}
\usepackage{xcolor}

\usepackage{graphicx} 
\usepackage{bbding}
\usepackage{hyperref}

\newcommand{\hlambda}{\hat{\lambda}}
\newcommand{\hP}{\hat{P}}
\newcommand{\tlambda}{\tilde{\lambda}}
\newcommand{\tP}{\tilde{P}}
\newcommand{\thlambda}{\bar{\lambda}}
\newcommand{\Ai}{\text{Ai}}

\newcommand{\dd}{\text{d}}

\newcommand{\ee}{\text{e}}

\newcommand{\eps}{\varepsilon}

\usepackage{comment}

\begin{document}
\title{Dynamics of an outlier in the Gaussian Unitary Ensemble}

\author{John Mateus}
\affiliation{Departamento de Física, Universidad de los Andes, Bogotá 111711, Colombia}

\author{Gabriel Téllez}
\affiliation{Departamento de Física, Universidad de los Andes, Bogotá 111711, Colombia}

\author{Frédéric van Wijland}
\affiliation{Laboratoire Mati\`ere et Syst\`emes Complexes, Université Paris Cité  \& CNRS (UMR 7057), 75013 Paris, France}
\affiliation{Yukawa Institute for Theoretical Physics, Kyoto University, Kyoto, 606-8502, Japan}

\begin{abstract}
We endow the elements of a random matrix drawn from the Gaussian Unitary Ensemble with a Dyson Brownian motion dynamics. We initialize the dynamics of the eigenvalues with all of them lumped at the origin, but one outlier. We solve the dynamics exactly which gives us a window on the dynamical scaling behavior at and around the Baik-Ben Arous-Péché transition. Amusingly, while the statics is well-known and accessible via the Hikami-Brézin integrals, our approach for the dynamics is explicitly based on the use of orthogonal polynomials.
\end{abstract}

\maketitle

\section{Introduction}
When a unit rank matrix is added to an otherwise random matrix of the most usual ensembles, the spectrum of the resulting random matrix can either retain some information on the rank-1 matrix, or completely wash out any information on the original perturbation. The transition between the two regimes, which occurs for asymptotically large matrices, is controlled by the amplitude of the rank-1 perturbation. It was first identified by \cite{edwards1976eigenvalue} but it is the work of \cite{baik2005phase} that renewed the interest in this BBP transition named after the authors Baik-Ben Arous-Péché of \cite{baik2005phase}. In the recent past, this transition has been shown to be at play in neural networks and information processing~\cite{saade2014spectral,sarao2019afraid,sarao2020marvels,pesce2022subspace} or in theoretical ecology \cite{fraboul2023artificial}, to cite but a few. The methods to explore the transition in the Gaussian Unitary Ensemble (GUE) case which we shall be interested in here are well-established. They are based on the Hikami-Brézin integrals~\cite{brezin1997extension, brezin1998level, brezin1998universal}.\\

In this work, what we bring is an approach of the BBP transition via a dynamical route, by endowing the matrix elements with a  Dyson Brownian motion dynamics, following the very pedagogical presentation of \cite{Biroli23}. By exactly solving the dynamics, we of course recover the celebrated static results, with the added value of obtaining purely dynamical, transient, results. Pictorially speaking with the {\it tour de France} in mind, we can see, as time elapses, how the peloton of eigenvalues manages to catch up, or not, the front runner. Interestingly, while Hikami and Brézin~\cite{brezin1997extension} explain that orthogonal polynomial-based methods are ineffective in their static approach, our dynamical approach specifically rests on such orthogonal polynomials.\\

This work begins with a brief description of the Dyson Brownian motion for the matrix elements, and of how, in the GUE, the dynamics can be expressed in terms of free fermions. Then, at fixed matrix size and time,  we solve the dynamics for an initial state in which all the eigenvalues a lumped together around the origin, but one outlier sitting a finite distance away. The final section explores the large matrix size and large time limits, where the Airy statistics of the largest eigenvalue is found right at the phase transition point.

\section{Dyson Brownian Motion}
For our introduction, we repeat the presentation of the BBP transition in dynamical terms as done in \cite{Biroli23}. At $t=0$ we start with an $N\times N$ rank one matrix $M(0)=\lambda_0 \mathbf{v} \mathbf{v}^T$ where $\mathbf{v}$ is a unit
vector and $\lambda_0$ the initial eigenvalue. The matrix $M(t)$ evolves
according to the equation
\begin{equation}
    \mathbf{M}(t+\dd t)=\mathbf{M}(t)+\mathbf{\dd g}(t),
\end{equation}
where $\mathbf{\dd g}(t)$ is a random matrix sampled from the Gaussian $\beta$-ensemble,
with $\beta=1$ for the Gaussian Orthogonal Ensemble (GOE), $\beta=2$ for the GUE, and $\beta=4$ for the Gaussian Symplectic Ensemble (GSE).
The matrix elements of $\mathbf{\dd g}$ are $\dd g_{ij}^\gamma(t)$, with
$(i,j)\in\{1,\cdots,N\}^2$, and $\gamma\in\{0,\ldots,\beta-1\}$ indexes the real
part ($\gamma=0$) and imaginary ($\gamma=1$) or quaternionic parts
($\gamma\in\{1,2,3\}$) of the matrix elements. The diagonal part of the matrix
$\mathbf{\dd g}$ is real and normal distributed $\dd g_{jj}^0(t)\sim
\mathcal{N}(0,\frac{2\dd t}{\beta N})$ and the off-diagonal elements are
distributed as $\dd g_{jk}^\gamma\sim\mathcal{N}(0,\frac{\dd t}{\beta N})$. After some
time $t$ has elapsed, the matrix $\mathbf{M}(t)$ has evolved to
\begin{equation}
    \mathbf{M}(t)=\lambda_0 \mathbf{v}\mathbf{v}^T+\mathbf{G}(t)
\end{equation}
with $\mathbf{G}(t)$ a random matrix from the Gaussian $\beta$-ensemble with
variance $\frac{t}{\beta N}$ for its off-diagonal elements. Standard techniques~\cite{sutherland1972exact,beenakker1994random} lead to the well-known Dyson Brownian motion for the $N$ eigenvalues $\lambda_i(t)$ of the matrix $\mathbf{M}(t)$:
\begin{equation}
    \frac{\dd\lambda_j}{\dd t} = \frac{1}{N}\sum_{l\neq j} \frac{1}{\lambda_j(t)-\lambda_l(t)} + 
    \sqrt{\frac{2}{\beta N}} \xi_j(t),\label{eq:DysonBrownianModel}
\end{equation}
where the independent Gaussian white noises $\xi_j$ have correlations $\langle
\xi_j(t)\xi_k(t')\rangle=\delta_{jk}\delta(t-t')$.

For technical reasons, it proves useful to change
the scaling of the eigenvalues by defining $\hlambda_j(t)=\sqrt{N}
\lambda_j(t)$. The equation of motion for a rescaled eigenvalue is
\begin{equation}
    \label{eq:langevin-hlambda}
    \frac{\dd\hlambda_j}{\dd t} = \sum_{l\neq j} \frac{1}{\hlambda_j(t)-\hlambda_l(t)} + 
    \sqrt{\frac{2}{\beta}} \xi_j(t).
\end{equation}
While the bulk of the eigenvalues $\lambda_j(t)$ evolve following a Wigner
semicircle law confined in $[-2\sqrt{t},2\sqrt{t}]$, the bulk of the
rescaled eigenvalues $\hlambda_j(t)$ follow a Wigner semicircle law confined in
$[-2\sqrt{tN},2\sqrt{tN}]$.

The Fokker-Planck equation for probability density function $\hP(\hlambda_1,\ldots,\hlambda_N,t)$ for the eigenvalues $\hlambda_j$ corresponding to the
Langevin equation~\eqref{eq:langevin-hlambda} is
\begin{equation}
    \label{eq:FP}
    \frac{\partial \hP}{\partial t} = 
    -\sum_{j=1}^N \frac{\partial}{\partial \hlambda_j} \left[\sum_{l\neq j} \frac{1}{\hlambda_j-\hlambda_l} \hP \right]
    + \beta^{-1}\sum_j \frac{\partial^2 \hP}{\partial \hlambda_j^2}.
\end{equation}
We apply a signed version of the Darboux transformation~\cite{spohn1987interacting,van1992stochastic} to transform this equation into its autoadjoint form by defining $\psi$ as
\begin{equation}
    \label{eq:psi-P}
    \hP(\hlambda_1,\ldots,\hlambda_N,t)= \Delta_N^{\beta/2}\psi(\hlambda_1,\ldots,\hlambda_N,t),
\end{equation}
where 
\begin{equation}
    \Delta_N= \Delta_N(\hlambda_1,\ldots,\hlambda_N)
\end{equation}
is the Vandermonde determinant
\begin{equation}\begin{split}
    \label{eq:vandermonde}
    \Delta_N(\hlambda_1,\ldots,\hlambda_N) =& \prod_{1\leq j < l\leq N} (\hlambda_l-\hlambda_j)\\
    =& \det\left(p_{j-1}(\hlambda_k)\right)_{1\leq j,k\leq N},
\end{split}\end{equation}
with $p_{n}(x)=x^{n}+\cdots$ is any monic polynomial of degree $n$. Later on, we
will use $p_n = 2^{-n} H_n$ where $H_n$ are the Hermite polynomials.

The autoadjoint form of Eq.~\eqref{eq:FP} is
\begin{equation}
    \label{eq:psi-HI}
    \frac{\partial\psi}{\partial t}=
    \frac{1}{\beta}\sum_{j=1}^N \frac{\partial^2 \psi}{\partial\hlambda_j^2} - H_I \psi 
\end{equation}
where 
\begin{equation}
    H_I = - \frac{1}{2}\left(1-\frac{\beta}{2}\right)
    \sum_{j=1}^N \sum_{k\neq j}\frac{1}{(\hlambda_j-\hlambda_k)^2}.
\end{equation}
Note that for $\beta=2$, the interaction term $H_I$ vanishes and the
transformation~\eqref{eq:psi-P} reduces to 
\begin{equation}
    \label{eq:psi-P-beta2}
    \hP(\hlambda_1,\ldots,\hlambda_N,t)= \Delta_N(\hlambda_1,\ldots,\hlambda_N)\psi(\hlambda_1,\ldots,\hlambda_N,t).
\end{equation}
Since the Vandermonde determinant $\Delta_N(\lambda_1,\ldots,\lambda_N)$ is
antisymmetric in its arguments while $\hP$ is symmetric, then the function
$\psi$ needs to be antisymmetric. In a quantum mechanical analogy, the function $\psi$ can be viewed as the wave function of identical fermions. Therefore at $\beta=2$ the initial problem is transformed into a problem of $N$ non-interacting fermions.

\section{Solving the dynamics for $\beta=2$}
From now on, we fix $\beta=2$. Equation~\eqref{eq:psi-HI} becomes 
\begin{equation}
    \frac{\partial\psi}{\partial t} = \frac{1}{2}\sum_{j=1}^N \frac{\partial^2 \psi}{\partial\hlambda_j^2},
\end{equation}
whose antisymmetric solutions are given by antisymmetrized plane-waves
\begin{equation}\begin{split}
    \psi(\hlambda_1,\ldots,\hlambda_N,t) =& \int \frac{\dd k_1}{2\pi}\cdots \int \frac{\dd k_N}{2\pi}
    c_{k_1\cdots k_N}\\&\times\frac{1}{N!} \det\left(\ee^{ik_l\hlambda_j }\right)_{1\leq j,l\leq N}
    \ee^{-\sum_{j=1}^N k_j^2 t/2}.
\end{split}\end{equation} 
The coefficients $c_{k_1\cdots k_N}$ are determined by the initial condition
\begin{equation}\begin{split}
    c_{k_1\cdots k_N} =& \frac{1}{N!}\int d\hlambda_1\cdots d\hlambda_N \det\left(\ee^{-ik_l\hlambda_j }\right)_{1\leq j,l\leq N}\\
    &\times
     \psi(\hlambda_1,\ldots,\hlambda_N,0).
\end{split}\end{equation}
At $t=0$, there is one eigenvalue $\hlambda_1^{0}=\hlambda_0$ from the
initial rank-one matrix $M(0)$ and the other eigenvalues are clustered around
zero $\hlambda_j^0=\eps_j\to 0$ for $j\geq 2$ where each $\eps_j$ is infinitesimally small. The nonzero $\eps_j$'s ensure a proper regularization on the $\hlambda_j^0$'s for the initial stages of the dynamics to be well defined. Indeed, in the Langevin equation for each eigevalue, the pairwise interaction force term $1/(\hlambda_j-\hlambda_k)$ is well defined only for $\hlambda_j\neq\hlambda_k$. The $\eps_j\to 0$ limit will be taken in due time. The initial condition for $\hP$ is
\begin{equation}
    \hP(\hlambda_1,\ldots,\hlambda_N,0) = 
    \frac{1}{N!} \sum_{\sigma\in S_N} 
    \prod_{j=1}^N \delta(\hlambda_j-\hlambda_{\sigma_j}^0),
\end{equation} 
where we symmetrized $\hP$ by summing over the $N!$ permutations $\sigma$ from the
permutation group $S_N$ of $N$ elements. Using Eq.~\eqref{eq:psi-P-beta2}, the
initial condition for $\psi$ is found to be 
\begin{equation}
    \psi(\hlambda_1,\ldots,\hlambda_N,0) =
    \frac{1}{N!} \frac{\det(\delta(\hlambda_j-\hlambda_k^{0}))_{1\leq j,k\leq N}}
    {\Delta_N(\hlambda_1^{0},\ldots,\hlambda_N^{0})}.
\end{equation}
With this we find the coefficients $c_{k_1\cdots k_N}$, $\psi$, and finally $\hP$ 
\begin{align}
    \hP(\hlambda_1,\ldots,\hlambda_N,t) &=
    \frac{1}{(2\pi)^N (N!)^2} 
    \frac{\Delta_N(\hlambda_1,\ldots,\hlambda_N)}{\Delta_N(\hlambda_1^{0},\ldots,\hlambda_N^{0})}
    &
    \nonumber\\
    &\times
    \int \dd k_1\cdots \dd k_N
    \det\left(\ee^{-ik_l\lambda_j^{0} }\right)_{1\leq j,l\leq N}\nonumber\\
    &\times
    \det\left(\ee^{ik_l\lambda_j}\right)_{1\leq j,l\leq N}
    \ee^{-\sum_{j=1}^N k_j^2 t/2}.
\end{align}
By writing out explicitly the determinants in the above expression, the
integrals over the $k_j$'s can be performed explicitly~\cite{Pandey91} and we arrive at
\begin{equation}\begin{split}
    \hP(\hlambda_1,\ldots,\hlambda_N,t) =&
    \frac{1}{(2\pi)^N N!} 
    \frac{\Delta_N(\hlambda_1,\ldots,\hlambda_N)}{\Delta_N(\hlambda_1^{0},\ldots,\hlambda_N^{0})}\\&\times
    \det(f(\hlambda_l-\hlambda_j^0))_{1\leq j,l\leq N},
\end{split}\end{equation}
where $f(x)=\int  \ee^{ikx-k^2t/2}\,\dd k =\sqrt{\frac{2\pi}{t}}\ee^{-\frac{x^2}{2t}}$. 

At this stage it is natural to introduce an additional rescaling of the eigenvalues $\tlambda_j=\hlambda_j/\sqrt{2t}$. The corresponding probability density function
$\tP$ for the rescaled eigenvalues $\tlambda_j$ is given by
\begin{equation}
    \tP(\tlambda_1,\ldots,\tlambda_N,t) = \hP(\hlambda_1,\ldots,\hlambda_N,t) (2t)^{N/2},
\end{equation}
upon taking into account the Jacobian $(2t)^{N/2}$ of the transformation. We have 
\begin{equation}\begin{split}
    \tP(\tlambda_1,\ldots,\tlambda_N,t) =&
    \frac{1}{\pi^{N/2} N!} 
    \frac{\Delta_N(\tlambda_1,\ldots,\tlambda_N)}{\Delta_N(\tlambda_1^{0},\ldots,\tlambda_N^{0})}\\
    &\times
    \det(\ee^{-\left(\tlambda_l-\tlambda_j^{0}\right)^2})_{1\leq j,l\leq N}.
\end{split}\end{equation}
Up to this point, this result is general for any deterministic initial
distribution of the initial eigenvalues $\tlambda_1^0,\ldots,\tlambda_N^0$. In
the following we shall take the limit $\tlambda_j^{0}\to 0$ for $j\geq 2$ while keeping
$\tlambda_1^{0}=\tlambda_0\neq 0$ fixed. In this limit, the Vandermonde
determinant in the denominator becomes 
\begin{equation}
    \Delta_N(\tlambda_1^{0},\ldots,\tlambda_N^{0}) = (-\tlambda_1^{0})^{N-1}
    \Delta_{N-1}(\tlambda_2^{0},\ldots,\tlambda_N^{0}).
\end{equation}
The Vandermonde $\Delta_{N-1}(\tlambda_2^{0},\ldots,\tlambda_N^{0})$ vanishes in
that limit and it is of order $O((\lambda_j^0)^{N-2})$ for each vanishing
eigenvalue $j\geq 2$. Therefore we need to expand the determinant in the
numerator to the same order. For this, we recall the generating function of the Hermite polynomials:
\begin{equation}
    \label{eq:hermite-generating}
    \ee^{-\left(\tlambda_l-\tlambda_j^{0}\right)^2} = \ee^{-\lambda_l^2}
    \sum_{n=0}^\infty \frac{H_n(\tlambda_l)}{n!}(\tlambda_j^{0})^n,
\end{equation}
and we use this expansion in the columns $j=2,\ldots,N$ of the determinant, up to
order $N-2$ in $\tlambda_j^0$:
\begin{equation}\label{eq:lemapp}\begin{split}
    \det\!\left(\ee^{-\left(\tlambda_l-\tlambda_j^{0}\right)^2}\right)_{1\leq j,l\leq N}=\\
    \det\left(\!\ee^{-\left(\tlambda_l-\tlambda_1^{0}\right)^2}, 
    \left(\!\ee^{-\tlambda_l^2}
    \sum_{n=0}^{N-2} \frac{H_n(\tlambda_l)}{n!}(\tlambda_j^{0})^n
    \right)_{j=2,\ldots,N}
    \right)_{l=1,\ldots,N}\\
    +O((\lambda_j^0)^{N-1}).
\end{split}\end{equation}
The matrix inside the latter determinant has one column with elements $\ee^{-\left(\tlambda_l-\tlambda_1^{0}\right)^2}$, $l=1,\ldots,N$, while the $j=2,\ldots,N$ remaining columns are encoded in the $N\times(N-1)$ matrix with elements $\left(\!\ee^{-\tlambda_l^2}
    \sum_{n=0}^{N-2} \frac{H_n(\tlambda_l)}{n!}(\tlambda_j^{0})^n
    \right)_{l,j}$ ($l=1,\ldots,N)$ (we shall use this shortcut notation further down in our derivation). We apply the lemma from Appendix~\ref{appA} to the above determinant to find that the probability density
function $\tP$ becomes,  in the $\tlambda_j^{0}\to 0$  limit, and for $j\geq 2$,
\begin{equation}\begin{split}
\tP(\tlambda_1,\ldots,\tlambda_N,t) =
\frac{\pi^{-N/2} (\tlambda_1^{0})^{-(N-1)}}{N!} 
\\
\times\det\left(2^{-(n-1)}H_{n-1}(\tlambda_j)\right)_{1\leq j,n\leq N}
\\\times\det\left( 
        \left(
    \frac{\ee^{-\tlambda_l^2}}{n!}
    H_n(\tlambda_l)
        \right)_{n=0,\ldots,N-2},
    \ee^{-\left(\tlambda_l-\tlambda_1^{0}\right)^2}
    \right)_{l=1,\ldots,N},
\end{split}\end{equation}
where we have used the representation given in Eq.~\eqref{eq:vandermonde} of
$\Delta_N(\tlambda_1,\ldots,\tlambda_N)$ in terms of Hermite polynomials. Let us now introduce the normalized wavefunctions of the harmonic oscillator $\phi_n(x) =
c_n e^{-x^2/2}H_n(x)$ with $c_n = (2^n n! \sqrt{\pi})^{-1/2}$, such that $\int
\phi_n(x)\phi_m(x)\,dx=\delta_{nm}$. The probability density function $\tP$ can
be written as 
\begin{equation}
    \tP(\tlambda_1,\ldots,\tlambda_N,t) =
    \frac{1}{N!} \det(A) \det(B)
\end{equation}
with 
\begin{equation}
    A=\left(\phi_{n-1}(\tlambda_j)\right)_{1\leq j,n\leq N},
\end{equation}
and
\begin{equation}\begin{split}
    B=\left( 
        \left(
    \phi_{n-1}(\tlambda_l)
        \right)_{n=1,\ldots,N-1},
        (\tlambda_1^{0})^{1-N} (N-1)!\right.\\\left.
        c_{N-1} \ee^{\tlambda_l^2/2}
    \ee^{-\left(\tlambda_l-\tlambda_1^{0}\right)^2}
    \right)_{l=1,\ldots,N}.
    \end{split}
\end{equation}
Note that the only difference between matrices $A$ and $B$ is their last column. If they were equal their determinants would of course be the same and $\det(A)\det(A) = \det(A^{T}A)$ would yield
$\det(K^0(\tlambda_l,\tlambda_j))_{1\leq j,l\leq N}$ with $K^0$ the familiar kernel of
the stationary GUE:
\begin{equation}
\label{eq:K0}
\begin{split}
    K^0(\tlambda,\tlambda') =& \sum_{n=0}^{N-1} \phi_n(\tlambda)\phi_n(\tlambda')\\
    =&\sqrt{\frac{N}{2}} \frac{\phi_{N}(\tlambda)\phi_{N-1}(\tlambda')-\phi_{N}(\tlambda')\phi_{N-1}(\tlambda)}{\tlambda-\tlambda'}.
\end{split}\end{equation}
In the last column of $B$, we resort to the expansion of Eq.~\eqref{eq:hermite-generating}:
\begin{equation}
    (\tlambda_1^{0})^{1-N} \ee^{\lambda_l^2/2}
    \ee^{-\left(\tlambda_l-\tlambda_1^{0}\right)^2}
    = \sum_{m=0}^{\infty} \frac{(\tlambda_1^{0})^{m-N+1}}{m!\, c_m} \phi_m(\tlambda_l).
\end{equation}
Since $\tlambda_1^{0}$ does not vanish, in principle all the terms in the sum
have to be kept. However, when evaluating $\det B$, one can subtract each of the $N-1$
previous $j$-th columns of $B$ multiplied by
$\frac{(N-1)!\,c_{N-1}}{j!\,c_j}(\tlambda_1^{0})^{j-N+1}$ to the last column, to
reduce the sum to $m \geq N-1$. This yields
\begin{equation}\begin{split}
    \det B = \det \left( 
        \left(
    \phi_{n-1}(\tlambda_l)
        \right)_{n=1,\ldots,N-1},\right.\\\left.
       \phi_{N-1}(\tlambda_l)
       + \sum_{m=N}^\infty \frac{(N-1)!\, c_{N-1}}{m!\, c_m} (\tlambda_1^{0})^{m-N+1} \phi_m(\tlambda_l)
    \right)_{l=1,\ldots,N}.
\end{split}\end{equation}
With this, we have built a biorthogonal structure \cite{Forrester2010} between the columns of $A$ and those of $B$. Then, writing $\det(A)\det(B)=\det(A^{T}B)$ we find 
\begin{equation}
    \tP(\tlambda_1,\ldots,\tlambda_N,t) =
    \frac{1}{N!} \det(K(\tlambda_l,\tlambda_j))_{1\leq j,l\leq N},
\end{equation}
with 
\begin{equation}\begin{split}
    K(\tlambda,\tlambda') = K^0(\tlambda,\tlambda')\\ + \phi_{N-1}(\tlambda)
    \sum_{m=N}^\infty \frac{(N-1)!\, c_{N-1}}{m!\, c_m} (\tlambda_1^{0})^{m-N+1} \phi_m(\tlambda').
\end{split}\end{equation}
It is remarkable that this modified kernel still satisfies the crucial property 
\begin{equation}
    \int K(\tlambda_1,\tlambda_3)K(\tlambda_3,\tlambda_2)\,\dd\tlambda_3 = K(\tlambda_1,\tlambda_2),
\end{equation}
and 
\begin{equation}
    \int K(\tlambda,\tlambda)\, \dd\tlambda = N.
\end{equation}
With this property, one can show~\cite{Mehta}, that 
\begin{equation}\begin{split}
    \int \det(K(\tlambda_l,\tlambda_j))_{1\leq j,l\leq p}\,\dd\tlambda_p =\\ 
    (N-p +1) \det(K(\tlambda_l,\tlambda_j))_{1\leq j,l\leq p-1}.
\end{split}\end{equation}
This allows us to integrate $\tP$ over any arbitrary number of eigenvalues and to
obtain the density of $p$ eigenvalues as 
\begin{equation}\begin{split}
    \tilde{n}^{(p)}(\tlambda_1, \ldots, \tlambda_p) = &
    \frac{N!}{(N-p)!} \int  \dd\tlambda_{p+1}\cdots \dd\tlambda_N \\&\tP(\tlambda_1, \ldots, \tlambda_N,t)\\
    =&
    \det(K(\tlambda_l,\tlambda_j))_{1\leq j,l\leq p}.
\end{split}\end{equation}
In particular, the one-body density of eigenvalues is 
\begin{equation}
    \tilde{n}(\tlambda) = K(\tlambda,\tlambda).
\end{equation}
 The stage is now set for extracting the large time and large matrix size asymptotics of the one-body density.

\section{Asymptotic analysis of the eigenvalue density}
The one-body density of eigenvalues can be split as 
\begin{equation}
    \tilde{n}(\tlambda) =  \tilde{n}_{\text{bulk}}(\tlambda) + \tilde{n}_{\text{out}}(\tlambda),
\end{equation}
where the bulk density
\begin{equation}
    \tilde{n}_{\text{bulk}}(\tlambda) = K^0(\tlambda,\tlambda),
\end{equation} 
represents the bulk density of the eigenvalues, and 
\begin{equation}
    \tilde{n}_{\text{out}}(\tlambda) = \phi_{N-1}(\tlambda)
    \sum_{m=N}^\infty \frac{(N-1)!\, c_{N-1}}{m!\, c_m} (\tlambda_1^{0})^{m-N+1} \phi_m(\tlambda),
\end{equation}
which can be associated to the density of the outlier eigenvalue that started at time $t=0$ at position $\tlambda_1^{0}$.

\begin{figure}
    \centering
    \includegraphics[width=0.49\textwidth]{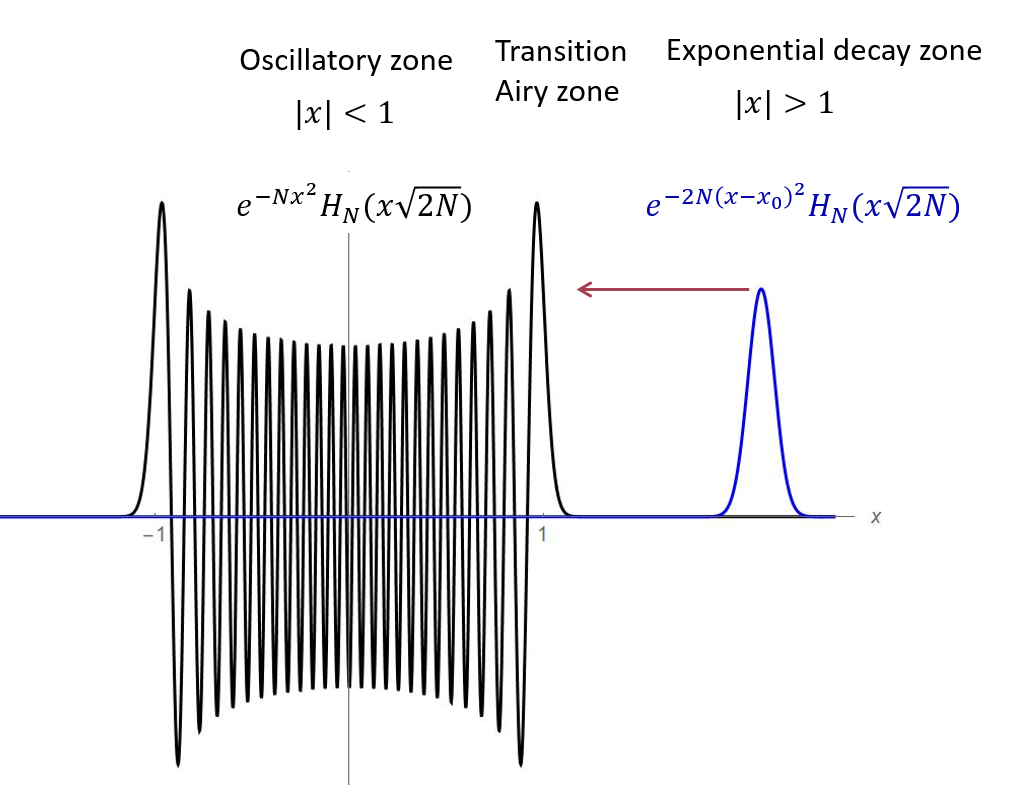}
    \caption{Illustration of the different regions in the Rotach--Plancherel asymptotics of the Hermite polynomials (for $N=50$). The black curve represents a typical term contributing to the bulk, whereas the blue curve shows the outlier contribution. Both are plotted as functions of $x = \lambda / \sqrt{4t}$. Each term has a different amplitude and has been rescaled by its respective maximum to allow comparison on the same plot. The outlier (in blue) is centered around $x^* = x_0 + (4x_0)^{-1}$, where $x_0 = \lambda_0 / \sqrt{4t}$ corresponds to the initial eigenvalue $\lambda_0$. As time $t$ goes by, $x^*$ moves from the exponential decay zone to the oscilatory zone, as indicated by the red arrow.}

    \label{fig:hermite-zones-outlier}
\end{figure}

Returning to the original variables $\lambda = \tlambda \sqrt{2t/N}$, the density
is 
\begin{equation}
    n(\lambda,t) = \sqrt{\frac{N}{2t}} \tilde{n}\left(\lambda\sqrt{\frac{N}{2t}}\right).
\end{equation}
From this, we see that in the $N\to\infty$ limit, we need to understand the asymptotic behavior of the Hermite polynomials $H_N(x\sqrt{2N})$ for large $N$ with $x=\lambda/\sqrt{4t}$ of order $1$ which are known as the Rotach-Plancherel asymptotics~\cite{Rotach25,Szego75,MenonLRMT}. These depend on the value of $x$. The region when $|x|<1$ is known as the oscillatory zone where the $N$ zeros of the Hermite polynomials are located. The region for $|x|>1$ is known as the exponential decay zone because in the function $\phi_N(x\sqrt{2N})$ the Gaussian factor creates an exponential decay. The transition between the two is given in terms of Airy functions. Figure \ref{fig:hermite-zones-outlier} shows these zones in a plot of the wave function $\phi_N$ along with the outlier contribution. In the following section, we will use these asymptotic expresions in the exponential decay zone and in the Airy zone which are given in Eqs. (\ref{eq:plancherel-rotach}) and (\ref{eq:plancherel-rotach-airy}). 

By applying the Christoffel–Darboux formula, Eq.~(\ref{eq:K0}), and taking the limit $\lambda' \to \lambda$, one obtains the following expression for the bulk density:
\begin{equation}
    \label{eq:bulk-density}
    \tilde{n}_{\text{bulk}}(\tlambda) = 
    N \phi_{N-1}(\tlambda)^2 - \sqrt{N(N-1)}\phi_{N}(\tlambda)\phi_{N-2}(\tlambda).
\end{equation}
In the $N\to\infty$  limit, the bulk density has a support in the oscillatory zone and it converges to the Wigner semicircle
law 
\begin{equation}
    \label{eq:wigner-semicircle}
    n_{\text{bulk}}(\lambda, t) \simeq 
    \begin{cases}
        \frac{N}{2\pi t} \sqrt{4t-\lambda^2},&\text{for }|\lambda| \leq \sqrt{4t},\\
        0&\text{for }|\lambda|>\sqrt{4t}.
    \end{cases}
\end{equation}
Thus, the bulk of the eigenvalues grows in an interval $[-2\sqrt{t},2\sqrt{t}]$,
while the outlier eigenvalue initially sits in the exponential decay zone and it is pushed by the repulsive interaction with the rest of the bulk eigenvalues. Indeed, at early times $t\ll 1$, in the region around the outlier $\lambda \sim \lambda_0$, we have $x=\lambda/\sqrt{4t}>1$ in the exponential decay zone. However, as time increases, $x$ decreases, and, at some time $t^*$, it will enter the oscillatory zone and blend with the bulk, see Fig.~\ref{fig:hermite-zones-outlier}. In the following subsections we study this behavior and transition in more detail.

\subsection{Early times evolution}

To better understand the behavior of the outlier density,
using Eq.~\eqref{eq:hermite-generating}, we further split it into two contributions
\begin{equation}
    \tilde{n}_{\text{out}}(\tlambda) = 
    \tilde{n}_{\text{out1}}(\tlambda) +
    \tilde{n}_{\text{out2}}(\tlambda)\, 
\end{equation} 
with
\begin{equation}
    \label{eq:nout1-prelim}
    \tilde{n}_{\text{out1}}(\tlambda) = 
    c_{N-1}^2 (N-1)! H_{N-1}(\tlambda) (\tlambda_1^{0})^{-(N-1)} \ee^{-(\tlambda-\tlambda_1^{0})^2},
\end{equation}
and 
\begin{equation}\begin{split}
    \label{eq:nout2-prelim}
    \tilde{n}_{\text{out2}}(\tlambda) = &
    c_{N-1}^2 (N-1)! H_{N-1}(\tlambda) (\tlambda_1^{0})^{-(N-1)}\\
    &\times
    \sum_{m=0}^{N-1} H_{m}(\tlambda) \frac{(\tlambda_1^{0})^{m}}{m!}.
\end{split}\end{equation}
At early times, the first term stands for the dominant contribution to the outlier density. 

Recalling that $\tlambda = \lambda \sqrt{N/2t}$, in the limit
$N\to\infty$, keeping $\lambda$ of order $O(1)$ and in the region outside the
bulk $\lambda^2>4t$, we expand the Hermite polynomials using the
Rotach-Plancherel asymptotics in the exponential decay zone
($x=\lambda/\sqrt{4t}>1$)~\cite{Rotach25,Szego75,MenonLRMT}:
\begin{equation}\begin{split}
    \label{eq:plancherel-rotach}
    H_{N+p}(x\sqrt{2N}) \sim &
    N^{\frac{N+p}{2}} 2^{\frac{N+p-1}{2}} \frac{(x+\sqrt{x^2-1})^{N+p+1/2}}{(x^2-1)^{1/4}}\\&\times
    \exp\left[
        \frac{N}{2} (x-\sqrt{x^2-1})^2
    \right].
    \end{split}
\end{equation}
 Note that, although the wave functions $\phi_N(x\sqrt{2N})$ decay exponentially fast when $x>1$, in $n_\text{out1}$ the Hermite polynomial is multiplied by a Gaussian centered around $\lambda_0$, instead of one centered around $0$ for the wave function. The Hermite polynomials increase algebraically fast in the region $x>1$, and this Gaussian will create a sort of ``bump'' in a region near $\lambda_0$ which is precisely the contribution of the outlier eigenvalue to the density. This is illustrated in Fig.~\ref{fig:hermite-zones-outlier}. This ``bump'' moves as time goes by, as we will show now. Using~\eqref{eq:plancherel-rotach}, we
find 
\begin{equation}
    n_{\text{out1}}(\lambda,t) \sim \sqrt{\frac{N}{2t \pi}}
    \frac{\lambda_0}{\lambda} 
    \frac{e^{-N f(\thlambda)}}{
        \left(1-\frac{4t}{\lambda^2}\right)^{1/4}
        \left(\frac{1}{2}\left(1+\sqrt{1-\frac{4t}{\lambda^2}}\right)\right)^{1/2}
    }
\end{equation}
where 
\begin{equation}\begin{split}
    f(\thlambda) =&
    \frac{(\thlambda - \thlambda_{0})^2}{2}  
    - \ln \frac{\lambda}{\lambda_{0}}
    - \frac{1}{2} \left(
        \frac{\thlambda}{2} - 
        \sqrt{\frac{\thlambda^2}{4}-1}
    \right)^2 \\&
    -\ln \left[
        \frac{1}{2}
        \left(
            1 + \sqrt{1-\frac{4}{\thlambda^2}}
        \right)
    \right]
\end{split}\end{equation}
with $\thlambda = \lambda/\sqrt{t}$. This can be approximated by a Gaussian
distribution using the Laplace method due to the exponential decay in $N$. The maximum of $f$ is obtained for
$\lambda=\lambda^*(t)$ which is a solution of $f'(\thlambda^*)=0$, namely
\begin{equation}
    \label{eq:outlier-th}
    \thlambda^* = \thlambda_0 + \frac{1}{\thlambda_0}.
\end{equation}
Returning to the original variables, this is
\begin{equation}
    \label{eq:outlier-eq}
    \lambda^*(t) = \lambda_0 + \frac{t}{\lambda_0},
\end{equation}
thus recovering the prediction made in \cite{Biroli23}. The variance can be deduced in the same spirit.
Expanding $f(\lambda)$ around $\lambda^*(t)$, we obtain
\begin{equation}
    n_{\text{out1}}(\lambda,t) \simeq \sqrt{\frac{N}{t \pi}}
    \lambda_0
    \frac{\exp\left[
        -\frac{N}{2} \frac{\lambda_0^2}{t(\lambda_0^2-t)}
        (\lambda - \lambda^*(t))^2 
    \right]}{
        \left(\lambda^2-4t\right)^{1/4}
        \left(\lambda^2+\sqrt{\lambda^2-4t}\right)^{1/2}
    }
\end{equation}
which is valid for $t<\lambda_0^2$ and in the region $\lambda > \sqrt{4t}$. If
we approximate in the prefactor $\lambda$ by $\lambda^*(t)$ (which is valid due to the large $N$ limit), we obtain a
Gaussian form for the outlier density:
\begin{equation}
    \label{eq:nout-guassian}
    n_{\text{out1}}(\lambda,t) \simeq \frac{1}{\sqrt{2\pi \sigma_\lambda(t)^2}}
    \exp\left[-\frac{(\lambda - \lambda^*(t))^2}{2\sigma_\lambda(t)^2}\right]
\end{equation}
with
\begin{equation}
    \label{eq:sigma-lambda}
    \sigma_\lambda(t)^2 = \frac{t(\lambda_0^2-t)}{N\lambda_0^2}.
\end{equation}
Ultimately, the bulk density, which grows as $2\sqrt{t}$ following the Wigner
semicircle law~(\ref{eq:wigner-semicircle}), will catch up with the outlier. This result is obtained from an exact calculation at $\beta=2$ valid for an arbitrary $N$, but at large $N$ the statistical law-of-large-numbers-based reasoning of \cite{Biroli23} extends to arbitrary $\beta$ as will be commented in the next section.
Using Eq.~\eqref{eq:outlier-eq}, we can find the time $t^*$ when the bulk catches up with the outlier corresponding to 
\begin{equation}
    \lambda^*(t^*) =
    2\sqrt{t^*}.\label{eq:reachtime}
\end{equation}
This corresponds to the time
\begin{equation}
    \label{eq:outlier-catch}
    t^* = \lambda_0^2 .
\end{equation}
For $t<t^*$ and in the region $|\lambda|>\sqrt{4t}$,
using the Plancherel-Rotach asymptotics in the exponential decay zone, one can
show that the contribution of $n_{\text{out2}}(\lambda)$ is negligible compared
to the first term $n_{\text{out1}}(\lambda)$.
After $t>t^{*}$, the contributions from $n_{\text{out}}(\lambda)$ cannot be
distinguished from the bulk density at leading order.

\begin{figure}
    \centering
    \includegraphics[width=0.49\textwidth]{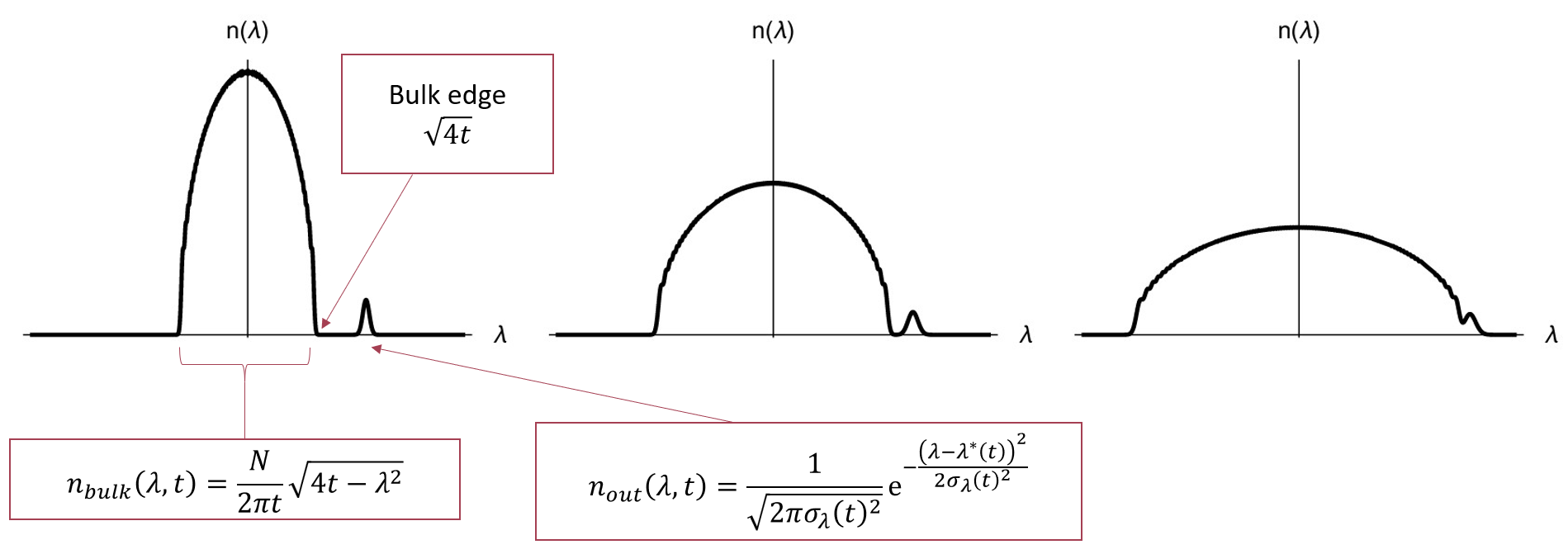}
    \caption{Snapshots of the eigenvalue density profile for increasing times from left to right.}
    \label{fig:bulk-outlier}
\end{figure}

The outlier density starts at $t=0$ as a delta peak $n_{\text{out}}(\lambda,0) =
\delta(\lambda-\lambda_0)$, then evolves as a drifting Gaussian centered around
$\lambda^*(t)$ given by Eq.~\eqref{eq:outlier-eq} up to time $t^*$, when the
bulk of eigenvalues catches up with the outlier. This is illustrated in Fig.~\ref{fig:bulk-outlier}

The variance
$\sigma_\lambda(t)^2$, given by Eq.~\eqref{eq:sigma-lambda}, is increasing from
time $t=0$ up to time $t=t^*/2=\lambda_0^2/2$, which is the time at which it reaches its maximum value
$\sigma_\lambda(t^*/2)^2 = \lambda_0^2/(4N)$. After this time $t^*/2$, the variance
decreases until $t=t^*$, and then $\sigma_\lambda(t^*)=0$. Note however that at
time $t=t^*/2$, $\lambda^*(t^*/2) = 1.5\lambda_0$, the bulk edge is at
$\sqrt{4t^*/2}=\sqrt{2}\lambda_0\simeq 1.41\lambda_0$, and
$\sigma_\lambda(t^*/2) = \lambda_0/(2\sqrt{N})$. Therefore, depending on $N$, the
decrease of the variance of the outlier, from $t=t^*/2$ to $t^*$, might be
masked by fluctuations of the edge of the bulk. This is illustrated further down in Sec.~\ref{sec:Num} where Eq.~\eqref{eq:outliervariancebetas} provides the reader with an alternative derivation of Eq.~\eqref{eq:sigma-lambda}. 

\subsection{Critical transition, beyond Gaussian statistics}

\begin{figure}[t]
  \centering
  \includegraphics[width=0.48\textwidth]{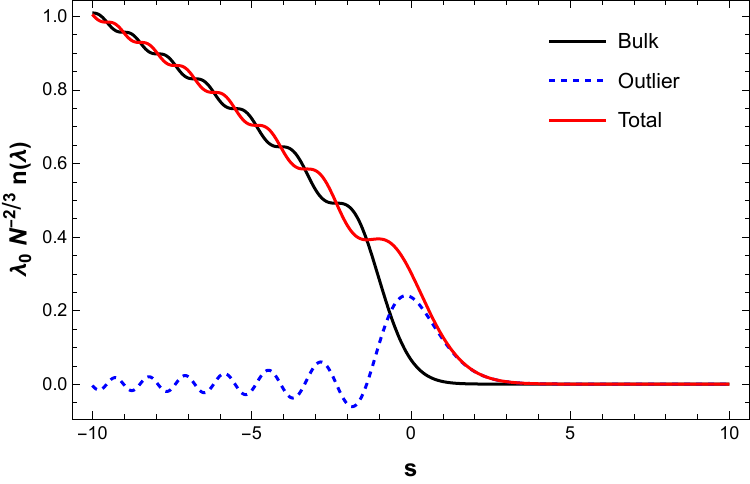}
  \caption{%
    Bulk contribution (black), outlier contribution (blue dashed),
    and total density (red) at the critical time $t^*$ in the Airy scaling zone. The variable $s$ denotes the rescaled position defined in Eq.~\eqref{eq:def-s}.
  }
  \label{fig:airy_density}
\end{figure}

We now study the density at the critical time $t^*$ when the bulk catches up
with the outlier. We are interested in the region around
$\lambda^*(t^*)=\sqrt{4t^*}=2\lambda_0$ at $t=t^*=\lambda_0^2$. All the terms in the density
contribute. Let 
\begin{equation}
    \label{eq:def-s}
    \lambda = 2\lambda_0 + \lambda_0 N^{-2/3} s,    
\end{equation}
with $s$ of order
$O(1)$. Using the results in \cite{Mehta, Forrester2010}, we can deduce that in terms of the scaling variable $s$  the bulk is described by an Airy scaling given
by
\begin{equation}
    n_{\text{bulk}}(\lambda,t^*) \sim 
    \frac{N^{2/3}}{\lambda_0}
    \left(\Ai'(s)^2 - s \Ai(s)^2 \right),
\end{equation}
where $\Ai(s)=\frac{1}{2i\pi}\int_{-i\infty}^{+i\infty} \ee^{z^3/3-zs}\dd z$ is the
Airy function. This asymptotic behavior is obtained by using the
Rotach-Plancherel asymptotics in the Airy transition zone 
\begin{equation}\begin{split}
    \label{eq:plancherel-rotach-airy}
    H_{N+p}( \sqrt{2N} + N^{-1/6}s/\sqrt{2}) \simeq&
    2^{\frac{N+p}{2}}(2\pi)^{1/4} 
    \sqrt{N!} N^{-\frac{1}{12}+\frac{p}{2}}\\&\times
    \ee^{N(1+N^{-2/3}s/2)^2}
    \Ai(s),
\end{split}\end{equation}
in Eq.~\eqref{eq:bulk-density}. Using Eq.~\eqref{eq:plancherel-rotach-airy} in
Eq.~\eqref{eq:nout1-prelim}, we obtain directly the first contribution to the
outlier density
\begin{equation}
\label{eq:nout1Airy}
    n_{\text{out1}}(\lambda) \simeq
    \frac{N^{2/3}}{\lambda_0} \Ai(s).
\end{equation}
For the second contribution, it is convenient to rewrite the sum over Hermite
polynomials in~(\ref{eq:nout2-prelim}) as 
\begin{equation}\begin{split}
    \label{eq:nout2-sum}
    S_N =& \sum_{m=0}^{N-1} 
    \frac{(\tlambda_1^{0})^m}{m!} H_m(\tlambda)\\
    =& \frac{(\tlambda_1^{0})^{N-1}}{(N-1)!}\int_0^{+\infty}
    \ee^{-t'}H_{N-1}\left(
        \tlambda + \frac{t'}{2\tlambda_1^{0}}
    \right)\, \dd t'.
\end{split}\end{equation}
This identity is obtained by using  
\begin{equation}
    H_m(x) = \frac{1}{\sqrt{\pi}}\int_{-\infty}^{+\infty}
    [2(x-i\tau)]^m \ee^{-\tau^2} \dd\tau,
\end{equation}
and
\begin{equation}
    \sum_{m=0}^{N-1} \frac{x^m}{m!}
    =
    \frac{1}{(N-1)!}\int_0^\infty \ee^{-t'} (x+t')^{N-1} \dd t'.
\end{equation}
In Eq.~(\ref{eq:nout2-sum}), we use the integral representation of the Hermite polynomial
\begin{equation}
    H_{N-1}(x)=\sqrt{\pi} \ee^{x^2} 
    \int_{-i\infty}^{+i\infty}
    z^{N-1} \ee^{-xz + z^2/4} \frac{\dd z}{2i\pi}.
\end{equation}
which allows us to write that
\begin{equation}\begin{split}
    \int_0^{+\infty}
    \ee^{-t'}H_{N-1}\left(
        \tlambda + \frac{t'}{2\tlambda_1^{0}}
    \right)\,\dd t'
    =\\ (2N)^{1+N/2}\sqrt{\pi}
    \int_0^{\infty} \dd u \int_{-i\infty}^{+i\infty} \frac{\dd z}{2i\pi}
    z^{-1} \ee^{2N g(z,u)}
    ,
\end{split}\end{equation}
with
\begin{equation}
    g(z,u)= \left(1+N^{-2/3}\frac{s}{2}+u-\frac{z}{2}\right)^2
    - u + \frac{1}{2}\ln z.
\end{equation}
The critical point of $g$, where $\partial_z g = 0$ and $\partial_u g = 0$, is
$z^*=1$ and $u=-\frac{s}{2N^{2/3}}$. We expand $g$ around this point, using
$g(z^*,u^{*})=\frac{1}{4}+N^{-2/3}s/2$, $\partial_{zu}^2 g(z^*,u^*)=-1$,
$\partial_{uu}^2 g(z^*,u^*)=2$. Let us note that $\partial_{zz}^2 g(z^*,u^*)=0$,
therefore it is necessary to go to the third order in the expansion with
$\partial_{zzz}^3 g(z^*,u^*)=1$. In terms of the variables $\tilde{z}=z-1$ and
$\tilde{u}= N^{2/3}u + s/2$, we have 
\begin{equation}
    2Ng(z,u) = 
    2Ng(z^*,u^*)
    - 2\tilde{u}\tilde{z}
    +\frac{1}{3}\tilde{z}^3
    + O(N^{-1/3}).
\end{equation}
With this, we obtain 
\begin{equation}
    S_N = \ee^{3N/2+N^{1/3}s}
    \int_{s}^{\infty} \Ai(u)\,\dd u
    \left(1+O(N^{-1/3})\right).
\end{equation}
Therefore, the second contribution to the density is
\begin{equation}
\label{eq:nout2Airy}
    n_{\text{out2}}(\lambda) \simeq
    -\frac{N^{2/3}}{\lambda_0} \Ai(s)
    \int_{s}^{\infty} \Ai(u)\,\dd u.
\end{equation}
Putting both contributions \eqref{eq:nout1Airy} and \eqref{eq:nout2Airy} together, and using the fact that
$\int_{-\infty}^{+\infty} \Ai(u)\,\dd u=1$, we obtain
\begin{equation}
    n_{\text{out}}(\lambda) \simeq
    \frac{N^{2/3}}{\lambda_0} \Ai(s)
    \int_{-\infty}^{s} \Ai(u)\,\dd u.
\end{equation}
In the slightly different context of the time-independent case, these Airy asymptotics for the modified kernel were also obtained in \cite{Forrester2010} (chapter 7). Figure \ref{fig:airy_density} shows the density $n(\lambda,t^*)$ in the Airy scaling near the edge, together with the bulk and the outlier contribution. Notice how the outlier presence makes a shift in the density oscillations of the bulk interchanging the maxima and minima with respect to the case when the outlier is not present. This is due to the natural repulsion between eigenvalues. 

At this stage we perform a numerical analysis of the stochastic equations of motion. The purpose is threefold: we use the exactly derived asymptotics to calibrate our simulations, then we probe the finite $N$ and finite $t$ asymptotics, and finally we explore the influence of the parameter $\beta$ for $\beta\neq 2$.

\section{Numerical explorations}\label{sec:Num}
\subsection{Implementing the numerical integration}
The Dyson Brownian motion dynamics of the eigenvalues can be performed by integrating Eq.~\eqref{eq:DysonBrownianModel} over an infinitesimal interval $\delta t$ according to a Euler scheme. The variation $\delta\lambda_j = \lambda_j(t+\delta t)-\lambda_j(t)$ is given by
\begin{equation}
    \delta\lambda_j = \frac{1}{N}\sum_{l\neq j}\frac{1}{\lambda_j(t)-\lambda_l(t)}\delta t+\sqrt{\frac{2}{\beta N}}\mathcal{N}_j(0,\delta t)\,,\label{eq:LangevinSimulation}
\end{equation}
where the $\mathcal{N}_j(0,\delta t)$, $j=1,\ldots,N$, are $N$ independent Gaussian variables with mean zero and variance $\delta t$. Thus, to first order in $\delta t$ we have
\begin{equation}
    \langle\delta\lambda_j\rangle = \frac{1}{N}\left\langle\sum_{l\neq j}\frac{1}{\lambda_j(t)-\lambda_l(t)}\right\rangle\delta t\,,\label{eq:MeanEigenStep}
\end{equation}
and
\begin{gather}
    \langle(\delta\lambda_j)^2\rangle = \frac{2}{\beta N}\delta t\,.\label{eq:NoiseSimulation}
\end{gather}
Using Eqs.~\eqref{eq:LangevinSimulation} and \eqref{eq:NoiseSimulation}, we performed a set of 1000 simulations (2000 in the case $\beta=2$), each with $N=100$ eigenvalues and an initial outlier $\lambda_0=0.1$ at time $t=0$, for values of $\beta \in [1.0, 5.0]$. The critial transition time is  $t^*=\lambda_0^2=10^{-2}$. Each simulation starts at $t=0$ and ends at $t=2t^*$, with 400-600 equidistant sample points taken for the eigenvalue time evolution. An adaptive time step $\delta t$ was used to evolve the system numerically, in order to prevent spurious jumps and unphysical crossings or exchanges of eigenvalues due to large repulsive forces when $\lambda_j$ approaches $\lambda_l$. To achieve this, we choose $\delta t$ by computing the average force and identifying the minimal eigenvalue spacing, $\Delta_{\text{min}}=\min{|\lambda_j-\lambda_l|,\, 1\leq j\neq l\leq N}$. Equation~\eqref{eq:MeanEigenStep} then provides the condition for choosing the time step:
\begin{equation}
    \delta t\leq N\Delta_{\text{min}}\left\langle\sum_{l\neq j}\frac{1}{\lambda_j(t)-\lambda_l(t)}\right\rangle^{-1}\sim10^x\,,\label{eq:timestep}
\end{equation}
here, $x<0$ denotes the order of magnitude of the time step $\delta t$. To ensure the validity of Eq.~\eqref{eq:timestep}, we set the simulation time step to $\delta t=10^{x-2}$, where $x$ is the solution of that equation. However, in some cases, $\delta t$ can become too large (say of the order $10^{-3}$) leading to a loss of resolution in tracking the eigenvalue trajectories. To avoid this, we impose a maximum allowed time step of $10^{-6}$ and take $\delta t=\min\{10^{x-2}\,, 10^{-6}\}$ during the simulations, depending on the situation.

The outcome of this procedure for the full eigenvalue and outlier dynamics is shown in Fig.~\ref{fig:simulation-evolution} for the case $\beta=2$ and in the video in the supplemental material \cite{sm:beta2_density_animation}. 
\begin{figure}[t!]
    \centering
    \includegraphics[width=0.49\textwidth]{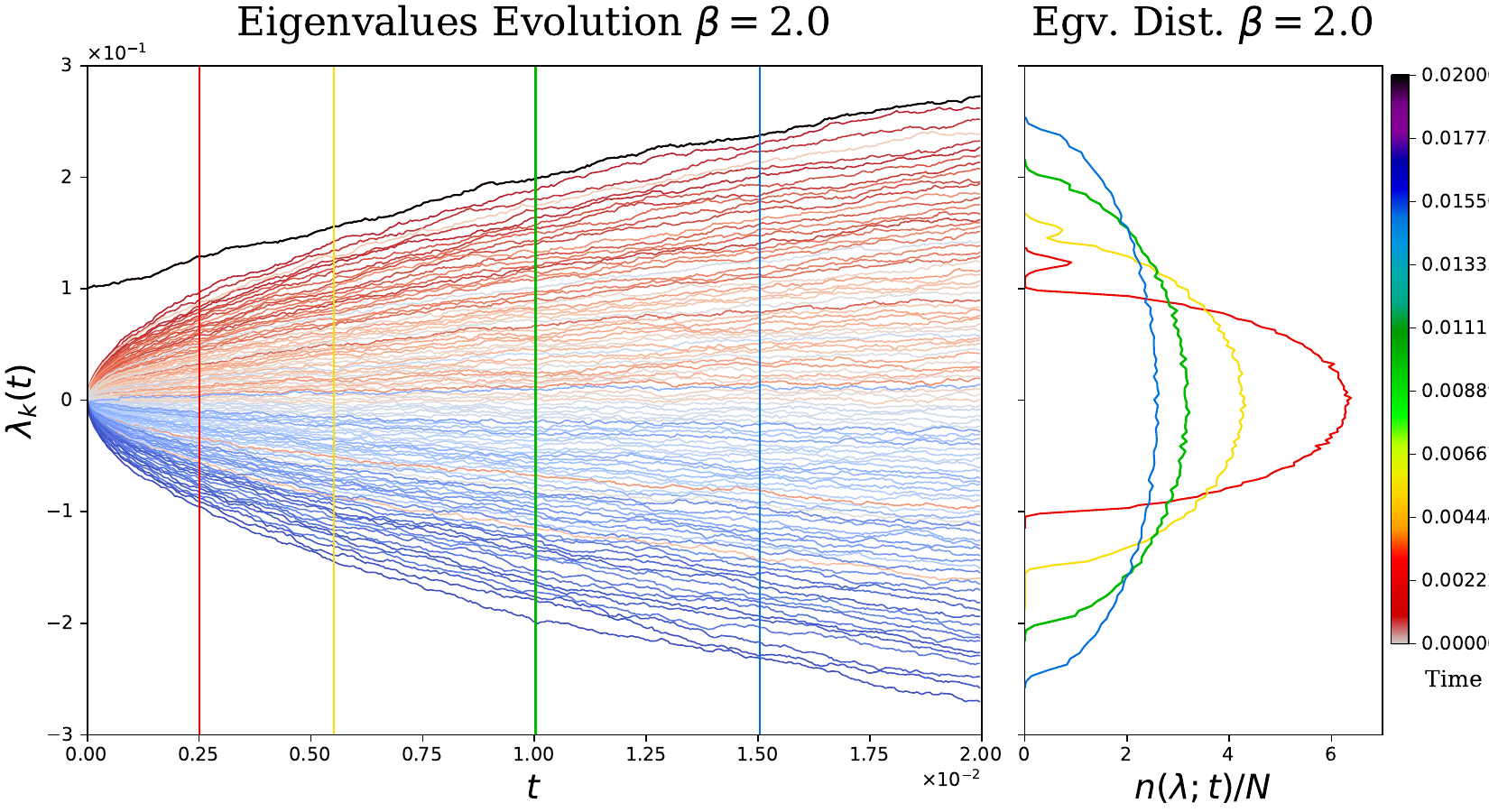}
    \caption{Time evolution from a single simulation (left) and the eigenvalue distribution aggregated over all simulations (right), using parameters $\beta=2$, $N=100$ eigenvalues, and $\lambda_0=0.1$ (shown as the black line in the left plot). Note that the eigenvalue trajectories do not intersect due to the adaptive time step defined in Eq. \eqref{eq:timestep}. Green line indicates the time $t^*$ given by Eq. \eqref{eq:outlier-catch}, when the bulk reaches the outlier. On the right, the probability distribution of the eigenvalues averaged over all simulations is shown at different times, that are marked with vertical lines in the left figure using the same color for each. Red and yellow lines highlight the separation between the bulk and the outlier for times $t<t^*$. As time progresses, the bulk catches up to the outlier (green line). After this point, the distinction between the bulk and the outlier becomes unclear (blue line).}\label{fig:simulation-evolution}
\end{figure}
On the left panel of Fig.~\ref{fig:simulation-evolution}, we see that the eigenvalue trajectories do not cross, which makes it possible to follow the evolution of each eigenvalue individually. On the right panel, the one-body probability distribution function of eigenvalues averaged over all simulations is shown at different times. For times $t<t^*$ (red and yellow lines) the peaks corresponding to the outlier can be distinguished from the bulk. As time progresses, the bulk approaches and eventually absorbs the outlier, as indicated by the green line at $t=t^*$. We now split our analysis between the benchmark  $\beta=2$ case and the $\beta\neq 2$ where little analytical progress is possible.

\subsection{Comparison with analytical predictions at $\beta=2$}
\begin{figure}[t]
    \centering
    \includegraphics[width=0.48\textwidth]{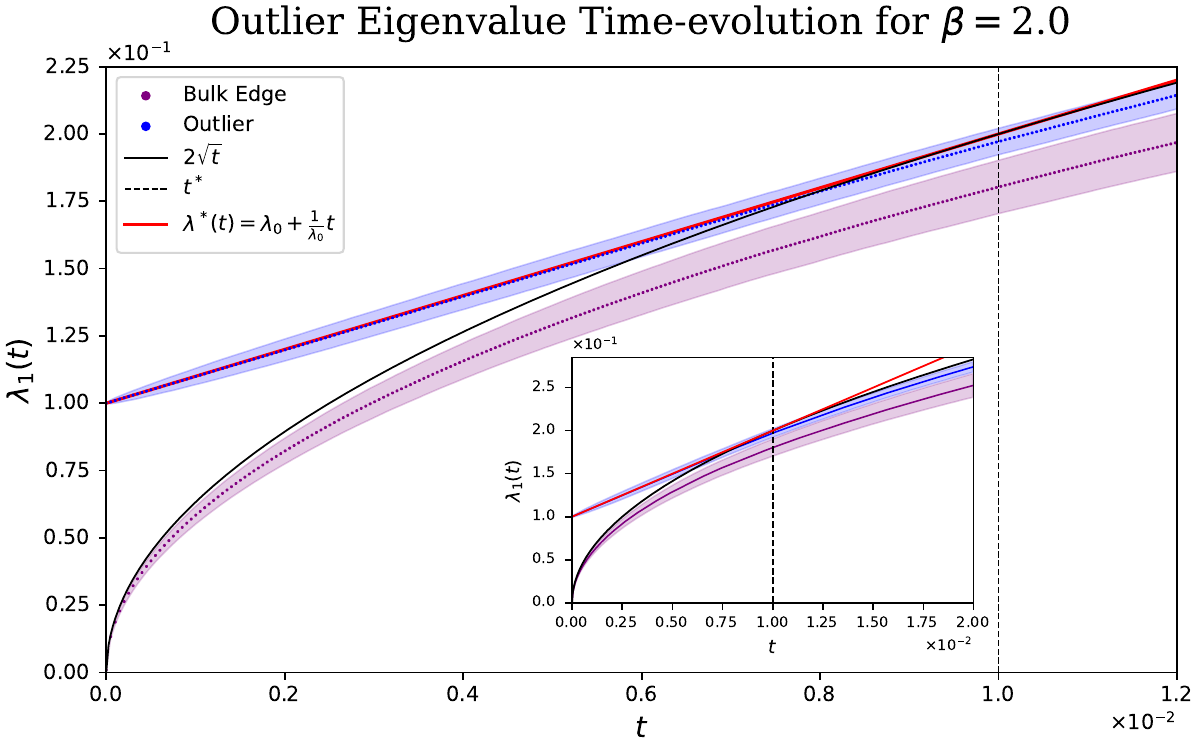}
    \caption{Evolution of the averaged outlier position (blue dots) for $N=100$ eigenvalues, $\beta=2$, and $\lambda_0=0.1$, compared with the theoretical prediction (red line) given by Eq.~(\ref{eq:outlier-eq}). Also shown is the evolution of the averaged edge of the bulk (purple dots), determined by tracking the eigenvalue closest to the outlier, averaged over all simulations. The bulk density predicted by the Wigner semicircle law is shown as a solid black line. Shaded areas represent the standard deviations of the simulated outlier and bulk positions, respectively. The inset displays the time evolution over the full simulation time domain.}\label{fig:outlier-evolution}
\end{figure}

Figure~\ref{fig:outlier-evolution} shows the time evolution of the averaged outlier (blue points) compared with the linear-in-time theoretical prediction (red line) of Eq.~\eqref{eq:outlier-eq}.

\begin{figure}[t]
    \centering
    \includegraphics[width=0.48\textwidth]{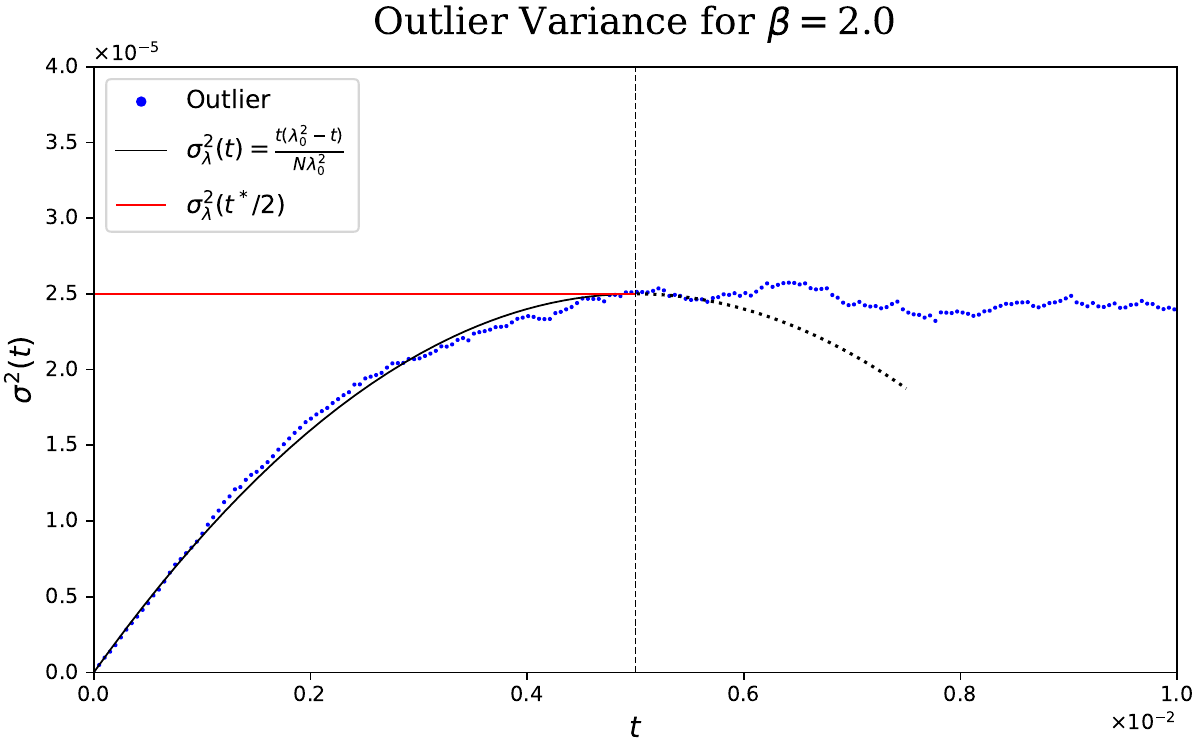}
    \caption{Time evolution of the outlier's variance for $\beta=2$ and $N=100$. At early times, Eq.~\eqref{eq:sigma-lambda} provides an excellent description. However, for $t>t^*/2$, fluctuations at the edge of the bulk mask the variance of the outlier.}\label{fig:squaredev-evolution}
\end{figure}
Using standard regression methods on the simulated averaged outlier data for $t\leq t^*$, we obtain a numerical slope of $m=9.7488\pm0.0071$, which is within 97.5\% of the theoretical value $m_{\text{theo}}=1/\lambda_0=10$. This demonstrates excellent agreement with the theoretical evolution of the outlier's peak density described by Eq.~(\ref{eq:outlier-eq}).

On the other hand, Eq.~\eqref{eq:sigma-lambda} predicts the behavior of the outlier’s variance at early times. As shown in Fig.~\ref{fig:squaredev-evolution}, for $t\ll t^{*}/2$, Eq.~\eqref{eq:sigma-lambda} provides an accurate description—at least for $\beta=2$—with the variance reaching its maximum at $t^{*}/2$ (indicated by the vertical dashed line). Beyond this point, bulk fluctuations dominate, masking the outlier’s variance evolution and rendering Eq.~\eqref{eq:sigma-lambda} invalid for $t > t^*/2$, as expected. We now turn to the fate of the exact $\beta=2$ results for $\beta\neq 2$.

\subsection{Conjectures at $\beta\neq 2$}
To address the $\beta\neq 2$ behavior, we first follow the approach of \cite{Biroli23}. Taking the large-$N$ limit in Eq.~\eqref{eq:DysonBrownianModel} for the outlier, we obtain:
\begin{gather}
    \frac{\dd\lambda_1}{\dd t} = \frac{1}{N}\int \dd\lambda\frac{n_{\text{bulk}}(\lambda,t)}{\lambda_1-\lambda}+\sqrt{\frac{2}{\beta N}}\xi_1(t)\,,\label{eq:OutlierLangevin}
\end{gather}
where $n_{\text{bulk}}(\lambda,t)$ is given in Eq.~\eqref{eq:wigner-semicircle}. Substituting this into Eq.~\eqref{eq:OutlierLangevin}, for the average value $\langle\lambda_1\rangle$ we find a differential equation
\begin{equation}
    \frac{\dd\langle\lambda_1\rangle}{\dd t} = \frac{\langle\lambda_1\rangle-\sqrt{\langle\lambda_1\rangle^2-4t}}{2t}\,,
\end{equation}
with solution
\begin{equation}
    \langle\lambda_1\rangle = \lambda_0+\frac{t}{\lambda_0}\,.\label{eq:meanlambda1}
\end{equation}
If we split $\lambda_1(t) = \langle \lambda_1 \rangle + \delta \lambda_1$  and use this in Eq.~\eqref{eq:OutlierLangevin} then, to first order in $\delta \lambda_1$ we obtain a Langevin equation for $\delta \lambda_1$ which, after using Eq.~\eqref{eq:meanlambda1}, is that of a Brownian bridge
\begin{equation}
    \frac{\dd\delta\lambda_1}{dt} = -\frac{\delta\lambda_1}{\lambda_0^2-t}+\sqrt{\frac{2}{\beta N}}\xi_1(t)\,.\label{eq:deltalambdaEq}
\end{equation}
Note that at $t=t^*=\lambda_0^2$, the ``velocity'' of the outlier $d\delta\lambda_1/dt$ diverges. This property survives in a generalized Dyson brownian motion model which includes many eigenvalues interactions and it is obtained from the evolution of two mutually free random matrices with arbitrary free cumulants \cite{BB2025}.

\begin{figure}[t]
    \centering
    \includegraphics[width=0.48\textwidth]{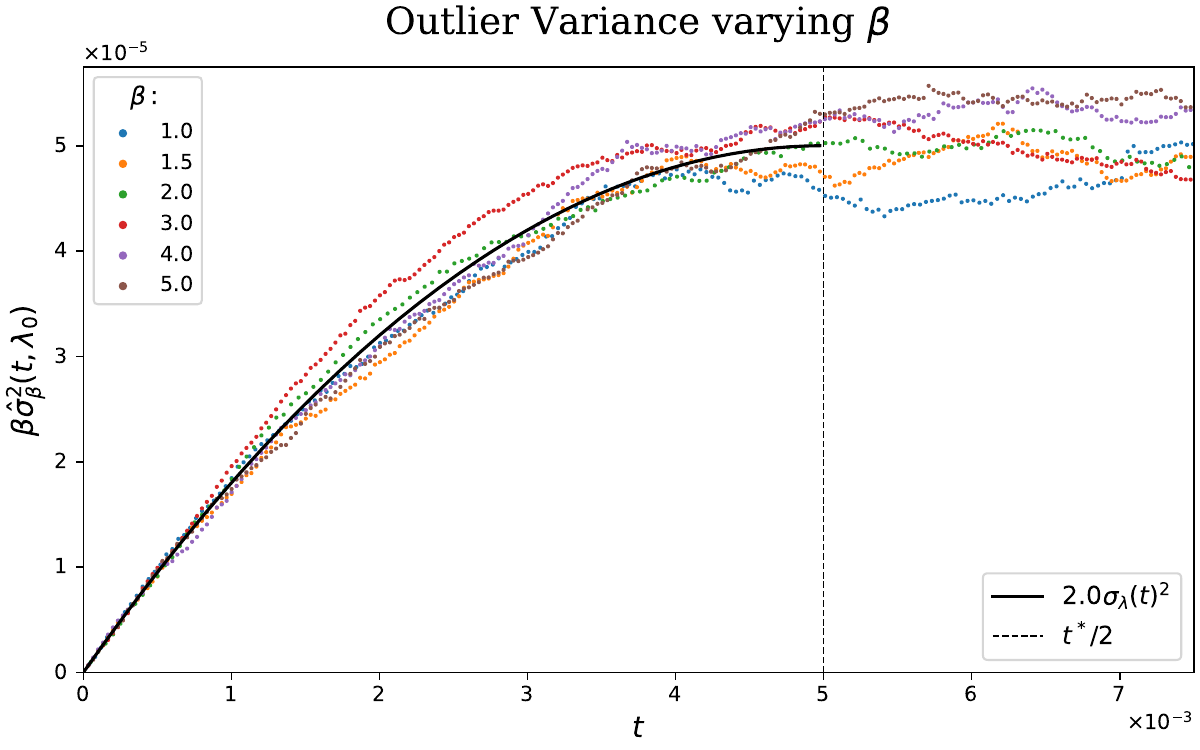}
    \caption{Time evolution of the outlier’s variance for various values of $\beta$. For early times $t\ll t^*/2$, Eq.~\eqref{eq:outliervariancebetas} provides an excellent description, with all curves collapsing onto a single one. The dashed line indicates $t^*/2$, while the solid black line represents the $\beta=2$ case computed using Eq.~\eqref{eq:sigma-lambda} with a prefactor of 2.0.}\label{fig:betasvariance-evolution}
\end{figure}

Returning to Eq.\eqref{eq:deltalambdaEq}, with a bit of It\=o calculus, the variance  $\hat\sigma^2_\beta(t,\lambda_0) = \langle(\delta\lambda_1)^2\rangle$ is found to evolve as
\begin{equation}
    \frac{\dd\hat\sigma^2_\beta(t,\lambda_0)}{\dd t} = -\frac{2\sigma^2_\beta(t,\lambda_0)}{\lambda_0^2-t}+\frac{2}{\beta N}\,,
\end{equation}
so that
\begin{equation}
    \hat\sigma^2_\beta(t,\lambda_0) = \frac{2t(\lambda_0^2-t)}{\beta N\lambda_0^2}\,.\label{eq:outliervariancebetas}
\end{equation}
Note that for $\beta=2$ we recover the result of Eq.~\eqref{eq:sigma-lambda} as we expected, thus, in terms of $\sigma_\lambda(t)^2$ the last equation can be written:
\begin{gather}
    \beta\hat\sigma^2_\beta(t,\lambda_0) = 2\sigma_\lambda(t)^2\,,
\end{gather}
This equality expresses that all curves  collapse onto a single one if we rescale the $y$-axis with the corresponding $\beta$ value for each simulation, as is indeed shown in Fig.~\ref{fig:betasvariance-evolution}.

\section{Conclusion}
Many of the formulas obtained in this work via a dynamical route echo those found in earlier works by Brézin and Hikami~\cite{brezin1997extension}. As they explain in their abstract ``The standard techniques, based on orthogonal polynomials,[...] are no longer available.'' It is rather amusing to see these very same orthogonal polynomials returning into action when using the dynamical route based on the Dyson Brownian motion. Our compact results rest on the observation that at $\beta=2$ the generator of the dynamics reduces, in the proper basis, to a free fermion Hamiltonian.
This allows to obtain exact results for arbitrary matrix sizes $N$, something that is not possible with other techniques for generic values of $\beta$ which only yield asymptotic results for $N\to\infty$. 
This suggests that obtaining results via perturbation theory around $\beta=2$ is within reach. Qualitatively we do not expect the dynamical scenario will change much (but the scaling functions will likely involve the Tracy-Widom distribution for the corresponding $\beta$). Among stimulating research directions, we would like to mention the possibility to explore via the dynamics other initial conditions  such as those studied in \cite{brezin1998universal}. It would be interesting to see how the more recently developed methods of Macroscopic Fluctuation Theory for random matrices~\cite{dandekar2024current, kgf2025} could provide an alternative route to study this problem.

\section*{Acknowledgments}
We are indebted to Giulio Biroli for directing us to several very relevant references. We acknowledge support from ECOS Nord
project C24P01 and Minciencias, Patrimonio Autónomo Fondo Nacional de
Financiamiento para la Ciencia, la Tecnología y la Innovación, Francisco José de
Caldas.
JM and GT acknowledge support of Universidad de los Andes, Facultad de
Ciencias, projects number INV-2024-199-3227 and INV-2023-176-2951, and its High
Performance Computing center (HPC).

\appendix

\section{Lemma}
\label{appA}
The lemma we use in Eq.~\eqref{eq:lemapp} in the main text can be phrased as follows. Consider the determinant $D$ defined by 
    \begin{equation}\begin{split}
    D=&\det\left(\ee^{-\left(\tlambda_l-\tlambda_1^{0}\right)^2},\right.\\ 
    &\left.\left( \ee^{-\tlambda_l^2}
    \sum_{n=0}^{N-2} \frac{H_n(\tlambda_l)}{n!}(\tlambda_j^{0})^n
    \right)_{j=2,\ldots,N}
    \right)_{l=1,\ldots,N},
 \end{split}   \end{equation}
then we can prove that
    \begin{equation}\begin{split}
        \label{eq:lemma}   
    D
    =& (-1)^{N-1}
    \Delta_{N-1}(\tlambda_2^{0},\ldots,\tlambda_N^{0})\\
    &
    \det\left( 
        \left(
    \frac{\ee^{-\tlambda_l^2}}{n!}
    H_n(\tlambda_l)
        \right)_{n=0,\ldots,N-2},
    \ee^{-\left(\tlambda_l-\tlambda_1^{0}\right)^2}
    \right)_{l=1,\ldots,N}.
  \end{split}  \end{equation}
The proof goes as follows. We begin by noting that $D$ is a polynomial of degree $N-2$ in each $\tlambda_j^0$
    ($j\geq2$) and has roots whenever $\tlambda_j^0=\tlambda_k^0$, because then
    columns $j$ and $k$ are identical and the determinant vanishes. Therefore
    $D$ is proportional to $\Delta_{N-1}(\tlambda_2^{0},\ldots,\tlambda_N^{0})$.
    To find the proportionality constant, we proceed by induction by looking at
    the coefficient of the highest power of $\tlambda_j^0$ in $D$. First, let us
    consider $D$ as a polynomial of $\tlambda_N^0$ of degree $N-2$. We notice
    that its roots are $\tlambda_2^0,\ldots,\tlambda_{N-1}^0$. Therefore 
    \begin{equation}
        D = C \prod_{j=2}^{N-1} (\tlambda_N^0-\tlambda_j^0),
    \end{equation}
    where $C$ is the coefficient of $(\tlambda_N^0)^{N-2}$ in $D$. To obtain
    $C$, note that $\lambda_N^0$ appears in the determinant only in the last
    column. If we keep only the highest power of $\tlambda_N^0$ in that last
    column, we obtain 
    \begin{equation}\begin{split}
        C =& \det\Bigg(
            \ee^{-\left(\tlambda_l-\tlambda_1^{0}\right)^2},\\
            &
    \left( \ee^{-\tlambda_l^2}
    \sum_{n=0}^{N-2} \frac{H_n(\tlambda_l)}{n!}(\tlambda_j^{0})^n
    \right)_{j=2,\ldots,N-1},\\
    &
    \ee^{-\tlambda_{l}^2} \frac{H_{N-2}(\tlambda_l)}{(N-2)!}
        \Bigg)_{l=1,\ldots,N}.
    \end{split}\end{equation}
    Now we subtract to each column $j=2,\ldots,N-1$ the last column multiplied
    by $(\tlambda_j^0)^{N-2}$, reducing the sum over $n$ to $n\leq N-3$, to obtain 
    \begin{equation}\begin{split}
        C = &\det\Bigg(
            \ee^{-\left(\tlambda_l-\tlambda_1^{0}\right)^2}, 
    \left( \ee^{-\tlambda_l^2}
    \sum_{n=0}^{N-3} \frac{H_n(\tlambda_l)}{n!}(\tlambda_j^{0})^n
    \right)_{j=2,\ldots,N-1},\\
    &
    \ee^{-\tlambda_{l}^2} \frac{H_{N-2}(\tlambda_l)}{(N-2)!}
        \Bigg)_{l=1,\ldots,N-1}.
    \end{split}\end{equation}
    We then repeat this same argument for column $N-1$ and so on till the
    second column to prove the lemma, Eq.~(\ref{eq:lemma}). The sign $(-1)^{N-1}$ arises from permuting the     first column to the last position by a cyclic permutation.  This concludes the proof of the lemma.

\bibliography{biblio}

\end{document}